\documentclass[10pt,leqno]{amsart}
\usepackage{graphicx}
\usepackage{indentfirst,csquotes}

\usepackage{amssymb,amsthm,amsmath}
\usepackage{xcolor,paralist,hyperref,titlesec,fancyhdr,etoolbox}

\titleformat{\section}[display]{\normalfont\huge\bfseries\centering}{\centering\chaptertitlename\thechapter}{10pt}{\Large}
\titlespacing*{\section}{0pt}{0ex}{0ex}

\hypersetup{ colorlinks=true, linkcolor=black, filecolor=black, urlcolor=black }

\usepackage{braket}
\usepackage{lipsum}
\usepackage{hyperref}
\begin{document}
\title{All-optical analogue of anti-skyrmions and anti-merons in difference-frequency-generation process} 
\author{Arannya Ghosh}
\author{Mukesh K. Shukla}
\author{Ritwick Das}

\date{\today}
\address{Optics and Photonics Centre, Indian Institute of Technology Delhi, Hauz Khas, Delhi 110016, India}
\email{arannya.ghosh@opc.iitd.ac.in}

\let\thefootnote\relax

\begin{abstract}
In the paraxial domain, a difference-frequency generation (DFG) process provides a versatile platform for exploring non-Hermitian dynamics in an all-classical interaction. This letter aims to show the existence of topologically non-trivial optical  analogues of anti-Skyrmionic states in an `anti' parity-time (anti-PT) symmetry-broken dynamical interaction in a DFG process. A practical DFG set-up, which is constituted by a periodically-poled lithium tantalate (PPLT) crystal and pumped by a visible (green) pump laser, is considered for ascertaining the non-Hermitian dynamics. In the anti-PT symmetry broken phase, a negatively phase-mismatched interaction gives rise to a topologically non-trivial pseudo-magnetization texture that could support anti-skyrmion-like quasi-particles in  signal and idler modes. The analysis also reveals the existence of anti-meron like pseudo-magnetization texture for a phase-matched DFG. 
\end{abstract} 

\bigskip
\maketitle
Topological quasi-particles, which have emerged as a major player in unifying diverse domains in core physical sciences, arise from \emph{symmetries} exhibited by vector-field configurations and display characteristics that are solely governed by topological invariants \cite{lei_topological_2025}. Skyrmion, a quasi-particle, was proposed to exist in a field configuration that exhibits non-trivial topology in real space and that resists backscattering during parallel transport \cite{skyrme_unified_1962}. By virtue of reliable and sophisticated material fabrication techniques, magnetic Skyrmions have captured significant attention in the field of spintronics \cite{zhang_magnetic_2023}. They provide versatile spintronic platforms for developing next-generation energy-efficient devices such as racetrack memory, logic devices, and neuromorphic computing components\cite{fert_magnetic_2017}. The impact of such breakthroughs and associated bottlenecks in the field of topological spintronics has triggered the search for Skyrmions in the domain of acoustics, electronics as well as optics \cite{ge_observation_2021,hlinka_skyrmions_2019,yang_optical_2025}. Optical Skyrmions open the possibility of scattering-free light transport in optical communication, image processing, and wide-area distributed sensor development \cite{cheng_navigating_2026}. In order to formulate a spintronic analogue, the role of `artificial gauge fields' in defining pseudo-spin basis is quintessential \cite{everschor-sitte_real-space_2014}. A natural extension of pseudo-spin basis in optics is the `polarization' ($s$ and $p$ basis) and the structural form of optical birefringence constitutes the 'artificial gauge fields'. In fact, this has manifested in the form of equivalent 'spin-orbit' interaction and topological studies in photonics\cite{wang_generation_2026}. Further, Skyrmionic textures such as Neel-type, Bloch-type and anti-type have been proposed using analogous form for magnetization in  interactions between different polarization basis \cite{lyu_topological_2025}. In a recent study, Y. Shen \emph{et al.} have presented analytical formalisms for generating a family of non-trivial magnetization textures, including Skyrmions, Merons, Bimerons, Hopfions etc. in paraxial polarization optics \cite{shen_generation_2022}. In addition to polarization, electromagnetic waves possess two more internal degrees of freedom (DoF), namely, momentum ($\vec{k}$) and energy ($\omega$), and their interaction with spatial beam structure is quite intriguing. It is worth noting that the aforementioned intrinsic DoFs make it more flexible to choose an appropriate basis which could advantageous from an application point of view. For example, Skyrmionic textures have been theoretically explored in a frequency upconversion process via creating topologically non-trivial pseudo-\emph{magnetization} in a non-centrosymmetric nonlinear crystal  \cite{karnieli_emulating_2021}. However, the constraints imposed on parameters for creating the complex pseudo-\emph{magnetization} texture are complicated and quite impractical in the context of experimental realization \cite{karnieli_emulating_2021,karnieli_experimental_2019}. A simpler route for inheriting a topological character is to adopt a Gaussian-pump-based frequency downconversion process in the paraxial domain. 
In this letter, we present an experimentally feasible anti-$\mathcal{PT}$ symmetry broken eigenvalue spectrum in a paraxial difference-frequency-generation (PDFG) process. Further, we show the existence of anti-Skyrmionic \emph{pseudo}-magnetization textures for a negative phase-mismatched PDFG configuration. In fact, we show analytically that a perfectly phase-matched PDFG process separates the two topological (trivial and non-trivial) phases of the non-Hermitian dynamics.

The interaction Hamiltonian ($\hat{H}$) for a PDFG process, under the undepleted pump approximation, exhibits a close resemblance to a quantum mechanical 'two-level' atomic system with signal and idler modes ($\ket{\omega_s}$ and $\ket{\omega_i}$) constituting the basis \cite{karnieli_all-optical_2018,westerberg_synthetic_2016}. Consequently, $\hat{H}= \frac{\vec{p}_T^2}{2M}-\vec{\sigma}.\vec{B}$ where, $\vec{B}=i|\kappa|(-\sin \phi \hat{x}+\cos{\phi}\hat{y})+\Delta k^{'} $ is an analogous representation of magnetic field and $\vec{\sigma}$ is the triad of Pauli spin matrices and $\phi$ is the pump beam phase\cite{westerberg_synthetic_2016,mondal_experimental_2022}. In this context, the analogous linear momentum $\vec{p}_t=-i\vec{\nabla}_t$ where $\vec{\nabla}_t$ is the transverse gradient operator that accounts for free-space diffraction. The nonlinear coupling coefficient $\vert \kappa \vert=\frac{4d_{eff}|A_p|}{n_s n_i}$ and phase-mismatch $\Delta k=k_p-k_s-k_i-\frac{2\pi}{\Lambda}$ are unambiguously defined with the grating period ($\Lambda$) gives the periodicity of periodic poling and $k_p,~k_s,~k_i$ are the wavevectors for the pump, signal and idler beams respectively that govern the phase-mismatch. Here, $d_{eff}$ and $|A_p|$ are the effective second-order nonlinear coefficient and pump beam amplitude respectively. For the sake of brevity, we define $\Delta k^{'}=\frac{\Delta k}{\sqrt{k_s k_i}}$ which is dimensionless \cite{karnieli_all-optical_2018}. By considering an effective mass operator $M=\frac{1}{m}(\sigma_z +\mathrm{I})$, where $m=\frac{2\sqrt{k_s k_i}}{k_s k_i}$ and $\mathrm{I}$ is the identity operator. The dressed state (\emph{i.e.} eigenstate of $\hat{H}$) is represented as $|\Psi\rangle=\begin{pmatrix}
    a_s^* & a_i
\end{pmatrix}^T$, where $a_s$ and $a_i$ are complex amplitudes of pseudo-spin basis $\ket{\omega_s}$ and $\ket{\omega_i}$ respectively. Using a transformation given by $a_s=\sqrt{k_i}\omega_s e^{i\frac{\Delta kz}{2}}a_s^{'}$ and $a_i=\sqrt{k_s}\omega_i e^{i\frac{\Delta kz}{2}}a_i^{'}$, the equation of motion could now be expressed as 
\cite{mondal_experimental_2022,nandi_all-optical_2025,nandi_all-optical_2026},

\begin{equation}
    -i\frac{\partial}{\partial {\Gamma}}\ket{\Psi'}=(\vec{\sigma}\cdot \vec{\Sigma)}\ket{\Psi'}
     \label{218}
\end{equation}
where $\ket{\Psi'}=\left(a_s^{'*}~~ a_i^{'} \right)^T$ and $\Sigma=(i|\kappa|\sin{\phi},-i|\kappa|\cos{\phi},\frac{\vec{p}_T^2}{2M}+\frac{\Delta k^{'}}{2})$ . $\Gamma=\sqrt{k_s k_i}z$ is the dimensionless evolution coordinate. We consider the Fourier transforms of the interacting fields \emph{i.e.} $A_j(x,y,z)=\iint_{-\infty}^{\infty}\tilde{A}_j(f_x,f_y,z) e^{2\pi i(f_xx+f_yy)}~df_x~df_y$ ($j \equiv p,s,i$) which transform the interaction Hamiltonian to \cite{kulkarni_classical_2022},
\begin{equation}
    \hat{\mathcal{H}}(f_x,f_y)=\begin{pmatrix}
        4\pi^2 f^2+\frac{\Delta k^{'}}{2} & -4\pi^2 \tilde{\kappa}^*~  \delta(f_x,f_y)  \\
         4\pi^2 \tilde{\kappa}~ \delta(f_x,f_y)  & -4\pi^2 f^2-\frac{\Delta k^{'}}{2}
    \end{pmatrix}
    \label{ham}
\end{equation}
where $f^2=\frac{(f_x^2+f_y^2)}{2M}$ ($f_x,~f_y$ are spatial frequencies along $x$ and $y$ directions respectively) is the dimensionless radial spatial frequency component, and $\delta(f_x,f_y)$ is the two-dimensional Dirac-delta function. In the parameter space constituted by $\Delta k'=\Delta k'(z)$ and $\tilde{\kappa}(z)$ ($\propto \vert \tilde{A}_p(f_x,f_y) \vert$), we consider any spatial dependence to conform to the adiabatic limit, which allows us to analyze the dynamics in the dressed-state picture \cite{allen_optical_2012}. The Hamiltonian ($\hat{\mathcal{H}}$) also allows us to ascertain any geometric manifestation via wavefront restructuring. It is worth noting that $\hat{\mathcal{H}}$ is non-Hermitian ( $\hat{\mathcal{H}}\neq \hat{\mathcal{H}}^\dagger $), which introduces singularities in the eigenvalue space. Since the spatial frequency (SF) space represents two-dimensional (2D) geometry, a radially symmetric pump beam manifests through circular symmetry of singular points, which are termed as 'exceptional rings' (ER). In the present context, the interaction dynamics in SF space serve as a pragmatic route for exploring the non-Hermitian topology of such nonlinear interaction process. In case of $\delta(f_x,f_y)=1$, depicting identical spatial frequencies for idler and signal beams, the Hamiltonian $\hat{\mathcal{H}}$ obeys anti-$\mathcal{PT}$ symmetry, \emph{i.e.} $(\mathcal{PT})\hat{\mathcal{H}}(\mathcal{PT})^{-1}=-\hat{\mathcal{H}}$ where $\mathcal{P} \equiv \sigma_x$ represents the parity operator ($\sigma_x$ being $x$-component of Pauli's spin matrices), while $\mathcal{T}$ denotes the anti-linear operator that transforms a function to its complex conjugate \cite{maamache_anti-pt_2020}. The eigenvalues of $\hat{\mathcal{H}}$ are,
\begin{equation}
    \lambda_\pm = \pm\sqrt{\left(4\pi^2 f^2+\frac{\Delta k^{'}}{2}\right)^2 -16\pi^4 |\tilde{\kappa}|^2 }
    \label{eigval}
\end{equation}
It is apparent that the eigenvalues are real when $f^2 > (|\tilde{\kappa}|- \frac{\Delta k^{'}}{8\pi^2})$ (for $\Delta k'\geq 0$) and imaginary for all other regions in SF space. While for $\Delta k'<0$ and $\vert \Delta k'\vert >8\pi^2 |\tilde{\kappa}|$, $\lambda_\pm$ are imaginary for $( - |\tilde{\kappa}|- \frac{\Delta k^{'}}{8\pi^2})<f^2 < (|\tilde{\kappa}|- \frac{\Delta k^{'}}{8\pi^2})$ and real for all values of $f^2$. According to the definition of anti-$\mathcal{PT}$ symmetry, the area in the SF space characterized by an imaginary eigenvalue spectrum that depicts preserved anti-$\mathcal{PT}$ symmetry, whereas a real spectrum indicates a region where anti-$\mathcal{PT}$ symmetry is broken \cite{bender_real_1998-1}. In this context, the anti-$\mathcal{PT}$ symmetry-broken regime exhibits conservative (or cyclical) dynamics that allow a geometric interpretation of the interaction mechanism. Further, the right and left eigenvectors (of $\hat{\mathcal{H}}$) are $\ket{\xi_\pm}=\begin{pmatrix}
        {4\pi^2 \tilde{\kappa}^*}&
        ({4\pi^2 f^2+\frac{\Delta k^{'}}{2}\mp \lambda_+})
    \end{pmatrix}^T$ and $\bra{\zeta_\pm }=\begin{pmatrix}
       4\pi^2 \tilde{\kappa} & (\pm\lambda_+ - 4\pi^2 f^2-\frac{\Delta k^{'}}{2})
   \end{pmatrix} $ respectively. The biorthogonal character of eigenstates could be verified via noting the fact that   $\bra{\zeta_\pm}\xi_\mp\rangle=0$ and $\bra{\zeta_\pm}\xi_\pm\rangle\neq0$. 
\begin{figure}[h]
    \centering
    \includegraphics[width=0.9\linewidth]{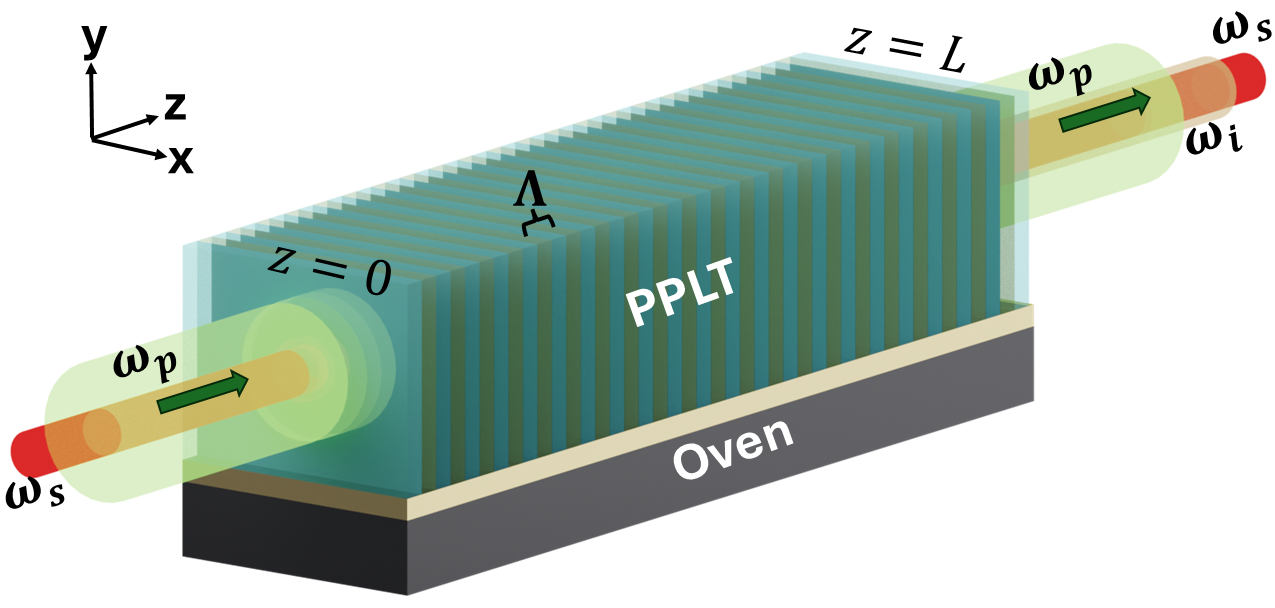}
    \caption{A schematic of difference-frequency generation (DFG) in PPLT crystal with a poling period $\Lambda$. The pump (green) and signal (red) beams are incident normally at $z = 0$ for the PPLT crystal while the idler beam (brown) is generated and exits at $z = L$. The pump beam is spatially broad in comparison with signal and idler beams.}
    \label{fig:placeholder}
\end{figure}
In order to analyze through an experimental framework, we consider a PDFG experimental configuration using a $30-mm$-long ($L$) periodically-poled lithium tantalate (PPLT) crystal where the pump and signal wavelengths are $\lambda_p = 532~nm$ and $\lambda_s = 820~nm$ respectively, and that results in an idler wavelength $\lambda_i = 1515~nm$. The PPLT crystal is assumed to be housed in an oven whose temperature could be continuously varied from room temperature (say $30^oC$) to $200^oC$. For a given grating period ($\Lambda$) of PPLT, the oven temperature is adjusted to achieve phase-matching ($\Delta k' = 0$) for the frequency conversion process. 
\begin{figure}[h]
    \centering
    \includegraphics[width=0.7\linewidth]{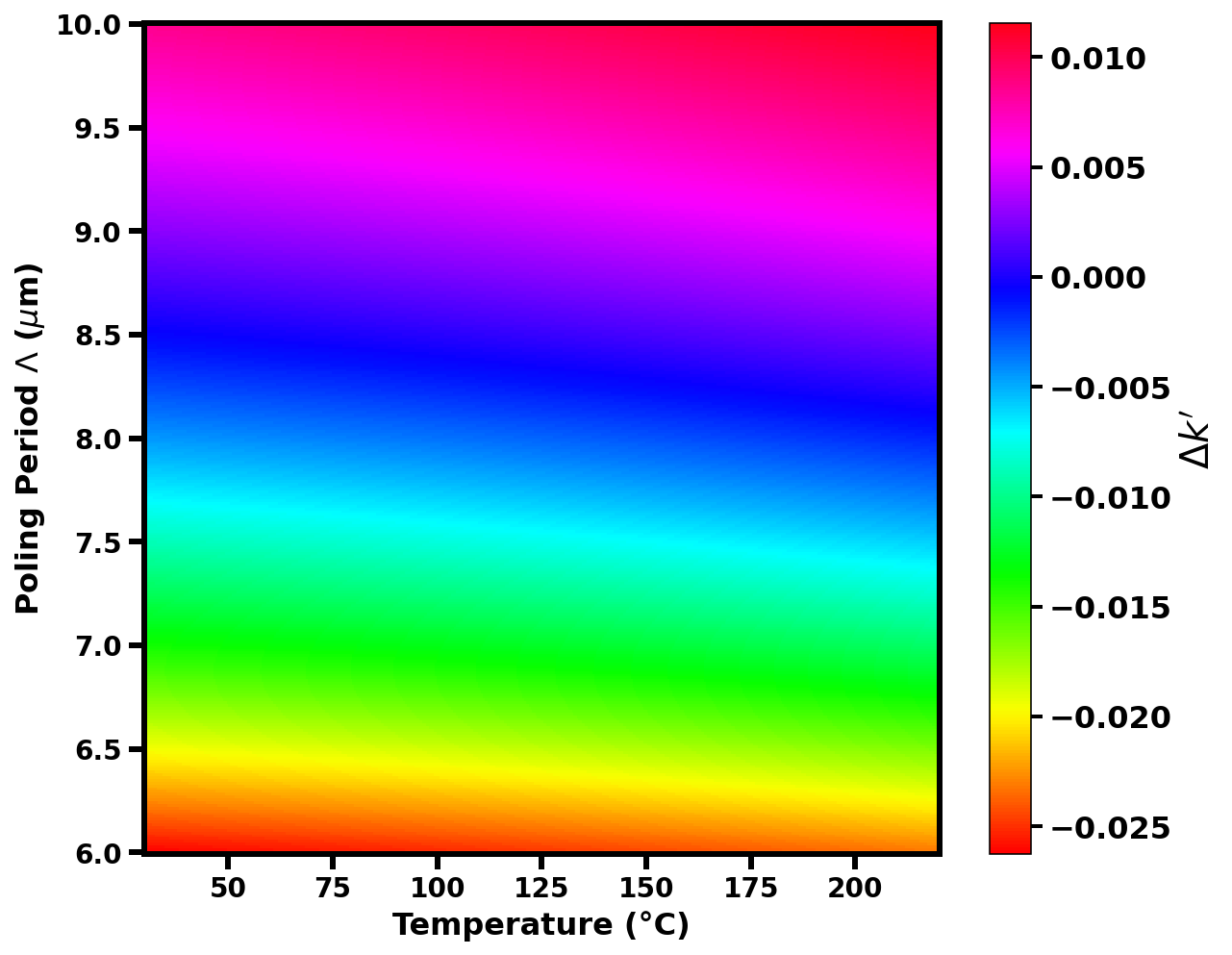}
    \caption{shows the variation of $\Delta k'$ as a function of poling period $(\Lambda)$ of PPLT crystal and the crystal oven temperature.}
    \label{fig:delta_k}
\end{figure}

\begin{figure}[h]
    \centering
    \includegraphics[width=1\linewidth]{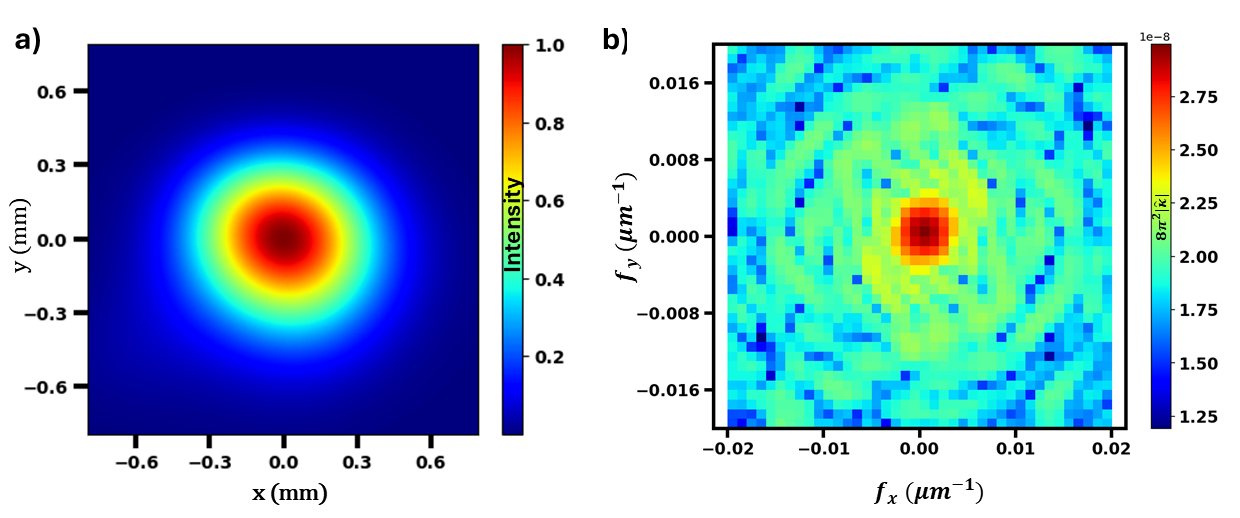}
    \caption{(a) shows the incident pump beam profile $x-y$ coordinates. (b) shows the 2D field distribution of $8\pi^2\vert \tilde{\kappa} (f_x,f_y)\vert$ in SF space. }
    \label{fig:eigvals}
\end{figure}
\begin{figure*}[t]
    \centering
    \includegraphics[width=1\textwidth]{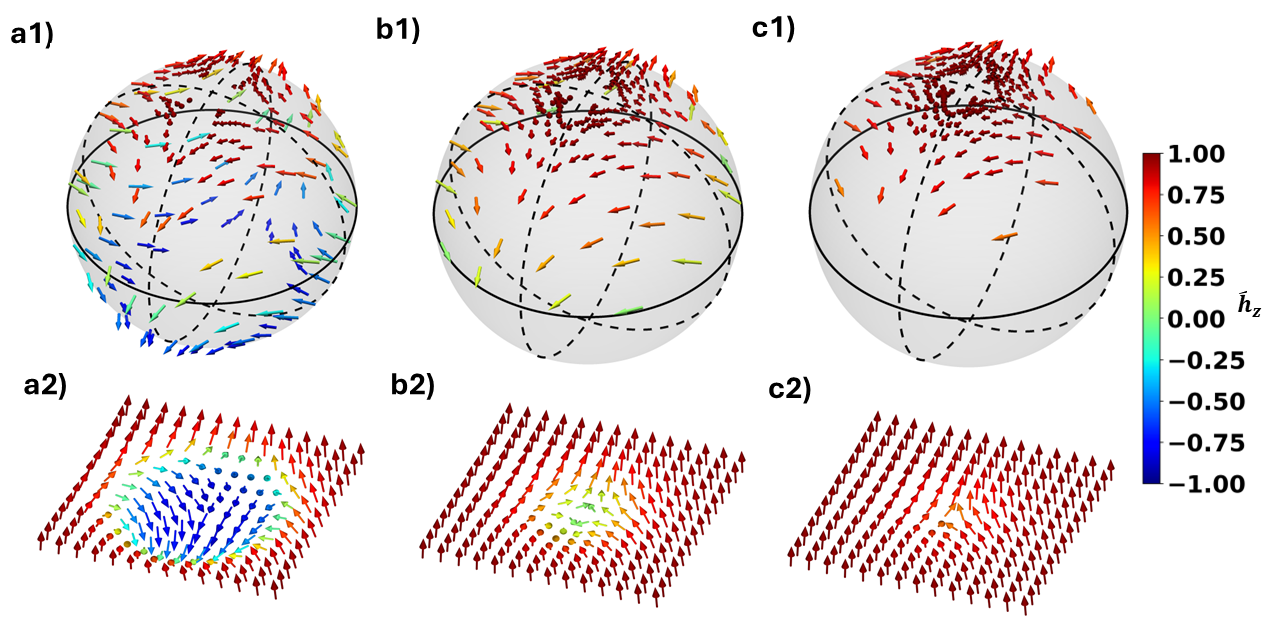}
    \caption{(a1),(b1) and (c1) shows the 3D vector field plot of pseudo-magnetization ($\hat{\mathbf{h}}(f_x,f_y)$) texture in the parameter space for $\Delta k'=-0.004$, $\Delta k' = 0$, and $\Delta k' = 0.002$, respectively. See visualizations. (a2), (b2) and (c2) are the stereographic (2D) projections for the 3D representations above.}
    \label{fig:3d_skyrmion}    
\end{figure*}
In the present context, however, we consider a variable grating PPLT (\emph{e.g.} fanned-out grating) crystal for exploring the variation in $\Delta k'$ as a function of grating period ($\Lambda$) at three different crystal oven temperatures ($T$). This defines positive and negative values of $\Delta k'$ that would be employed for obtaining a real (or imaginary) eigenvalue spectrum in SF space with reference to Eq. (3) (see Fig.(\ref{fig:delta_k})) \cite{bruner_temperature-dependent_2003}. From Fig.(\ref{fig:delta_k}), we choose three different values of phase-mismatch factor at $T = 200^oC$ (i)  $\Delta k'=0$ for $\Lambda = 8.22~\mu m$ (ii) $\Delta k'=-0.004$ for  $\Lambda = 7.77~\mu m$ and (iii) $\Delta k'=0.002$ for $\Lambda = 8.5~\mu m$ for further analysis. A pragmatic estimate of $\tilde{A}_p(f_x,f_y)$ is obtained from a recorded pump beam ($\lambda_p = 532~nm$) profile as shown in Fig. 3(a) at an optical average power of $\approx 100~mW$ (with $M^2 \approx 1.15$). The pump beam profile has been recorded using a CCD with pixel size $4.6~\mu m \times 4.6~\mu m$ (shown in Fig. \ref{fig:eigvals}a). By assuming a spatial Gaussian beam profile for the pump beam, we note that $|\tilde{\kappa}|$ varies between $1.25\times10^{-8}\leq8\pi^2\vert \tilde{\kappa}\vert\leq2.75\times10^{-8}$ across the SF plane (shown in figure-\ref{fig:eigvals}b). 


\begin{figure}[h]
    \centering
    \includegraphics[width=\linewidth]{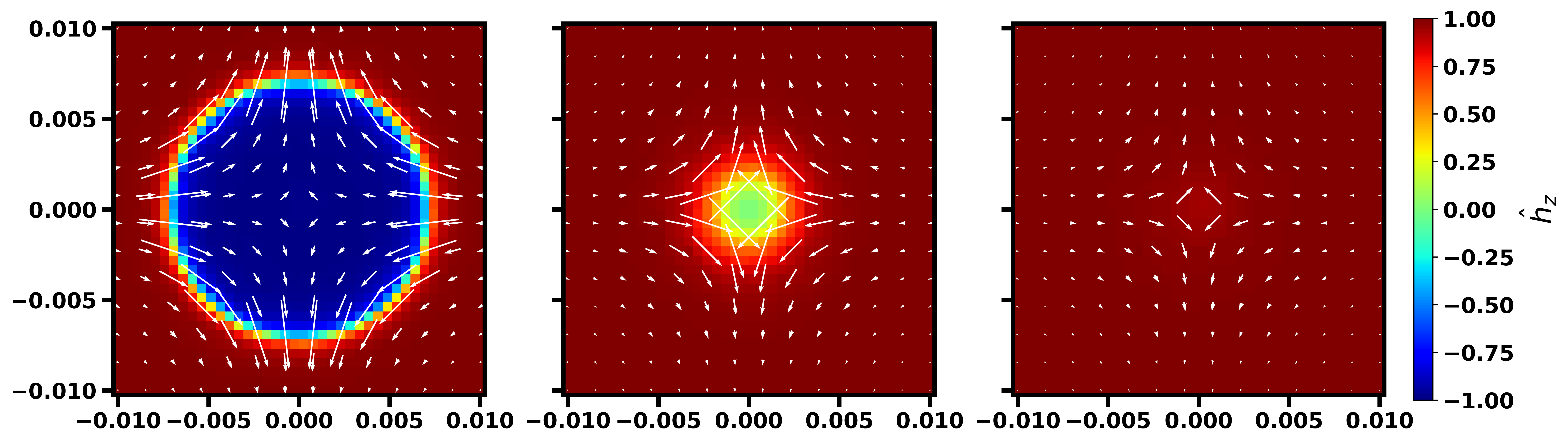}
    \caption{(a), (b) and (c) show the top-view of $2D$ vector field plots (stereographic projection) of $h_x$ and $h_y$ for $\Delta k'=-0.004$,$\Delta k' = 0$, and $\Delta k' = 0.002$, respectively.}
    \label{fig:skyrmion_vertical}
\end{figure}
We rephrase the Hamiltonian description in Eq.(\ref{ham}) via 
\begin{equation}
\tilde H(f,\phi)
=\begin{pmatrix}
    h_z & h_x+ih_y\\
    -h_x+ih_y & -h_z
\end{pmatrix} 
\end{equation}
where, $h_x = -|\tilde{\kappa}'({f})|\cos\phi,
h_y = |\tilde{\kappa}'({f})|\sin\phi,
h_z = 4\pi^2 f^2+\frac{\Delta k'}{2}$ constitute the pseudo-magnetization vector and $|\tilde{\kappa}'|=4\pi^2 |\tilde{\kappa}|$. Consequently, the Skyrmion number is defined in the SF space as \cite{lin_photonic_2023} ,
  \begin{equation}
N_{\mathrm{sk}}
=
\frac{1}{4\pi}
\int d^2 f\;
\hat{\mathbf h}\cdot
\left(
\partial_{f_x}\hat{\mathbf h}
\times
\partial_{f_y}\hat{\mathbf h}
\right).
\label{skyrmion_def}
\end{equation}
where the unit magnetization vector is $\hat{\mathbf h}=\frac{\vec{h}}{\vert h\vert} = \frac{1}{E}(-|\tilde{\kappa}'| \cos{\phi},|\tilde{\kappa}'| \sin{\phi},h_z)$ such that $\vert h\vert=\sqrt{|\tilde{\kappa}'|^2+
\left(4\pi^2 f^2+\frac{\Delta k'}{2}\right)^2 }= E $.
\eqref{skyrmion_def} could be represented in polar coordinates as, 
\begin{equation}
N_{\mathrm{sk}}
=
\frac{1}{4\pi}
\int_0^{\infty} df
\int_0^{2\pi} d\phi\;
\hat{\mathbf h}\cdot
\left(
\partial_f\hat{\mathbf h}
\times
\partial_\phi\hat{\mathbf h}
\right).
\label{skyrmion_2}
\end{equation}
which redefines the Skyrmion number as (see supplemental document),
\begin{equation}
    N_{sk}=\frac{1}{2}\left[\frac{h_z(f)}{\sqrt{|\tilde{\kappa}'(f)|^2+h_z(f)^2}}\right]_0^\infty
    \label{skyrmion_finalw}
\end{equation}
Since the pump is spatially Gaussian (see Fig. \ref{fig:eigvals},  $~|\tilde{\kappa}'(f)|\to0$ as $f \to \infty$ which, effectively, manifests as $N_{sk}\vert _{f \to \infty}=\frac{1}{2}$. On the other hand, $~N_{sk}=\frac{\Delta k'/2}{\sqrt{|\tilde{\kappa}'(0)|^2+(\Delta k'/2)^2}}$ when $f=0$ which, in the present context, is $\pm \frac{1}{2}$ depending on the positive and negative phase-mismatch ($\Delta k'$). Consequently, the Skyrmion number, which is a topological index, takes the form, 
\begin{equation}
    N_{sk}=\frac{1}{2}\left[1-sgn(\Delta k')\right]
    \label{skyrmion_short}
\end{equation}
When $\Delta k'\neq 0$, it is apparent from \eqref{skyrmion_short} that $N_{sk}=1$ and $N_{sk}=1/2$ for $\Delta k'<0$ and $\Delta k'=0$ respectively which depict topologically non-trivial pseudo-magnetization texture. On the other hand, $N_{sk}=0$ when $\Delta k' > 0$ which represents a topologically trivial interaction dynamics. To elucidate this point, we consider a PPLT-based DFG configuration discussed before. For $\Delta k' = -0.004$, we have presented 3D vector field plot of $\hat{\mathbf{h}}$ in Fig. (\ref{fig:3d_skyrmion}-a1) where $h_x$ and $h_y$ ($\propto |\tilde{A_p}(f)|$) are in-plane components, while $h_z$ is oriented perpendicular to that plane. In the SF plane, a negative $\Delta k'$ ($N_{sk} = 1$) ensures opposite orientations for $\hat{\mathbf{h}}$ at the north and south poles of the corresponding Bloch sphere which signifies a Skyrmionic pseudo-magnetization texture (see Visualization 1). For comparison, Fig. (\ref{fig:3d_skyrmion})-(b1) and (c1) are plots for $\hat{\mathbf{h}}$ in the parameter space when $\Delta k' = 0$ and $\Delta k' = +0.002$, respectively (see Visualization 2 and Visualization 3). In case of $\Delta k' = 0$, $N_{sk} = \frac{1}{2}$ as per \eqref{skyrmion_short} which implies non-trivial topology. However, $h_z$ remaining positive results in pseudo-magnetization ($\hat{\mathbf{h}}$) that is non-zero up to the equitorial plane in the 3D parameter space. Such pseudo-magnetization texture depicts the existence of topological quasi-particles known as 'merons'. In consistency with the description above, $\Delta k' > 0$ depicts trivial topology where $\hat{\mathbf{h}}$ remains confined to the northern hemisphere without reaching the equitorial plane. The stereographic projection in Fig. \ref{fig:3d_skyrmion} (a2) represents the 2D vector field, $\hat{\mathbf{h}}$ which flips through an adiabatic route while traversing from  distant points in the SF plane ($f_x,f_y \to\infty$) to the beam center $f_x,f_y \to 0$ when $\Delta k' < 0$. This flipping is not observed in the stereographic projections shown in Fig. \ref{fig:3d_skyrmion}(b2) ($\Delta k' = 0$) and (c2) ($\Delta k' = 0.002$). 
It is worth reiterating that the Skyrmion number, as defined in \eqref{skyrmion_def}, quantifies the geometric phase acquired by the state-vector after encircling the unit Bloch sphere. 
In the SF space, the Skyrmion number (defined in \eqref{skyrmion_2}), could be better articulated as the product of two integers \cite{shen_generation_2022}:
\begin{equation}
    N_{sk}=p \cdot  v
\end{equation}
where $p$ and $v$ define the polarity and vorticity respectively. In physical terms, the value of $p$ is dictated by the orientation of $\hat{\mathbf{h}}$ at the north pole and the south pole on Bloch sphere. For example, $p=1$ when $\hat{\mathbf{h}} = +\hat{z}$ at the North Pole ($f_x,f_y \to\infty$) as well as at the South Pole ($f_x,f_y \to 0$) of the Bloch sphere. On the other hand, $p = -1$ when $\hat{\mathbf{h}} = +\hat{z}$ at the North Pole and $\hat{\mathbf{h}} = -\hat{z}$ South Pole. 

Vorticity ($v$) is determined by the orientation of in-plane vectors $\vec{h}_T$ ($= {h}_x\hat{x}+h_y\hat{y}$) at the equitorial plane in the parameter space. In general, $v=1$ depict the existence of Bloch-type or Neel-type Skyrmions whereas $v=-1$ gives rise to an anti-Skyrmionic magnetization texture \cite{shen_optical_2024}. In order to elucidate the relevance of aforementioned definitions, we present a vector plot on  2D spatial frequency ($f_x-f_y$) plane in Fig. \ref{fig:skyrmion_vertical}(a)-(c). The existence of a saddle point in all three cases ($\Delta k'=-0.004$, $\Delta k'=0$ and $\Delta k'=+0.002$) depicts $v = -1$, a signature that is governed by the Hamiltonian (\eqref{ham}). Therefore, the spatial evolution of signal and idler beams in a negatively phase-mismatched ($\Delta k' < 0$) DFG process pertains to $p = v = -1$ which gives rise to an anti-Skyrmionic pseudo-magnetization texture with $N_{sk} = 1$\cite{koshibae2016theory}. On the other hand, phase-matched ($\Delta k'= 0$) DFG results in an anti-meron like quasi-particle generation in the dressed-sate picture with $N_{sk} = 1/2$\cite{guo_meron_2020}. On the other hand, a positive phase-mismatched ($\Delta k' > 0$) is a topologically trivial dynamical interaction.
In conclusion, we explored the DFG dynamics in a PPLT crystal with a Gaussian pump laser beam at $532~nm$ wavelength. Via mapping the dynamical variables in a DFG process onto that for a two-level non-Hermitian Hamiltonian, we show that the dynamics obeys anti-$\mathcal{PT}$ symmetric interaction between basis modes \emph{i.e.} signal and idler modes. In the broken anti-$\mathcal{PT}$ symmetry phase, we discover that the DFG process gives rise to a topologically non-trivial pseudo-magnetization texture which is quantified through a non-zero Skyrmion number. Subsequently, via using a quantitative analysis based on polarity and vorticity of the pseudo-magnetization texture, we show the existence of anti-Skyrmion quasi-particles in the negatively phase-mismatched DFG process which are present in the signal and idler mode-fields. The findings of this work aim to provide a nonlinear frequency-conversion based platform for exciting and manipulating non-Hermitian topological quasiparticles. Since the DFG process could be scaled on a broad wavelength range, the results pave the way for the development of anti-Skyrmion-encoded parametric devices that facilitate reliable spatial-mode manipulation and topologically safeguarded beam-shaping in nonlinear optics.


\bigskip

\bibliographystyle{unsrt}
\bibliography{ab_1}




\end{document}